\documentclass[prl,twocolumn,superscriptaddress,showpacs]{revtex4}
\usepackage[latin1]{inputenc}
\usepackage{amsmath}
\usepackage{amsfonts}
\usepackage{amssymb}
\usepackage{graphicx}
\usepackage{subfigure}
\usepackage{epstopdf}
\usepackage{epsfig}
\usepackage{multirow}

\begin{document}

\addtolength{\topmargin}{0.125in}
\addtolength{\oddsidemargin}{0.00in}

\pagestyle{empty}

\title{Magnetic fluctuations and spin freezing in non-superconducting LiFeAs derivatives}
\author{J.~D.~Wright}
\affiliation{Department of Physics, Clarendon Laboratory, University of Oxford, Parks Road, Oxford, OX1 3PU, United Kingdom }
\author{M.~J.~Pitcher}
\affiliation{Department of Chemistry, University of Liverpool, Crown Street, Liverpool, L69 7ZD, United Kingdom }
\author{W.~Trevelyan-Thomas}
\affiliation{Inorganic Chemistry Laboratory, University of Oxford, South Parks Road, Oxford, OX1 3QR, United Kingdom }
\author{T.~Lancaster}
\affiliation{Department of Physics, Durham University, South Road, Durham, DH1 3LE, United Kingdom}
\author{P.~J.~Baker}
\affiliation{ISIS Muon Facility, STFC Rutherford Appleton Laboratory, Harwell Oxford, Didcot, OX11 0QX, United Kingdom }
\author{F.~L.~Pratt}
\affiliation{ISIS Muon Facility, STFC Rutherford Appleton Laboratory, Harwell Oxford, Didcot, OX11 0QX, United Kingdom }
\author{S.~J.~Clarke}
\affiliation{Inorganic Chemistry Laboratory, University of Oxford, South Parks Road, Oxford, OX1 3QR, United Kingdom }
\author{S.~J.~Blundell}
\affiliation{Department of Physics, Clarendon Laboratory, University of Oxford, Parks Road, Oxford, OX1 3PU, United Kingdom }

\begin{abstract}
We present detailed magnetometry and muon-spin rotation data on polycrystalline samples of overdoped, non-superconducting LiFe$_{1-x}$Ni$_x$As ($x = 0.1,\,0.2$) and Li$_{1-y}$Fe$_{1+y}$As ($0\leq y\leq 0.04$) as well as superconducting LiFeAs. While LiFe$_{1-x}$Ni$_x$As exhibits weak antiferromagnetic fluctuations down to $1.5\,{\rm K}$, Li$_{1-y}$Fe$_{1+y}$As samples, which have a much smaller deviation from the $1:1:1$ stoichiometry, show a crossover from ferromagnetic to antiferromagnetic fluctuations on cooling and a freezing of dynamically fluctuating moments at low temperatures. We do not find any signatures of time-reversal symmetry breaking in stoichiometric LiFeAs that would support recent predictions of triplet pairing.
\end{abstract}  


\maketitle

Of all the known Fe-based superconductors, LiFeAs remains one of the most intriguing: unlike other pnictides, such as BaFe$_2$As$_2$ \cite{avci2012} and NaFeAs \cite{parker2010}, LiFeAs is a superconductor in its stoichiometric form \cite{pitcher2008, tapp2008} and any chemical substitution on the Fe-site (with Co or Ni for instance) causes a reduction in the transition temperature, $T_{\rm c}$ \cite{pitcher2010}. In contrast with other systems, no ordered magnetic phase or structural transition has yet been observed in LiFeAs, a fact that has provoked much debate given the tendency for band-structure calculations to predict similar magnetic ground states to those seen in other pnictides \cite{zhang2010, li-liu2009, li2009}. Applied pressure suppresses superconductivity, but does not induce magnetism \cite{mito2009}. 

Magnetic fluctuations, however, have been observed. Inelastic neutron scattering (INS) experiments uncover incommensurate fluctuations close to the wavevector $Q = (0.5, 0.5, 0)$ in both superconducting \cite{taylor2011, qureshi2012} and non-superconducting (apparently Li-deficient) \cite{wang2011} forms of LiFeAs. This $Q$-vector is the same as that which gives rise to the striped antiferromagnetic ground state seen in other pnictides and which is predicted by the aforementioned theoretical studies \cite{zhang2010, li-liu2009, li2009}. This suggests that there is a degree of commonality between LiFeAs and other pnictides but that some crucial difference prevents it from ordering magnetically as they do. ARPES measurements of the Fermi surface \cite{borisenko2010} suggest that this difference may be the comparatively poor nesting between electron and hole pockets.

To account for this, and to fit LiFeAs into a unified scheme for the pnictides, a recent report \cite{wang2012} suggested that LiFeAs behaves analogously to the electronically overdoped versions of other systems. Common features in INS data support this idea, and it would explain why further electron doping, such as with Co or Ni, only reduces $T_{\rm c}$. 
However, it would also suggest that removing electrons (hole doping) may induce an ordered magnetic state, and no evidence for this has yet been reported. An alternative approach to account for the special status of LiFeAs is centred around the suggestion that it may exhibit triplet pairing \cite{brydon2011, aperis2013}, which has gathered some experimental support \cite{baek2012, baek2013, hanke2012}.

In this paper we present studies of two series of non-superconducting LiFeAs derivatives, LiFe$_{1-x}$Ni$_x$As ($x = 0.1$ and $0.2$) and Li$_{1-y}$Fe$_{1+y}$As ($y = 0.01, 0.018$ and $0.04$), as well as a stoichiometric (superconducting) compound. Superconductivity is known to be suppressed when $x\geq 0.1$ and $y\geq 0.01$ \cite{pitcher2010}. For the Fe-rich series, the concentration of Fe on the Li site was obtained from a Rietveld refinement of the structure against both synchrotron x-ray and neutron powder diffraction data. All samples were found to be of very high purity: no extra phases were observed in the diffraction data and magnetisation data taken at room temperature ruled out the presence of any magnetic impurities the diffraction experiments may have missed. Details of these analyses, along with synthesis procedures, can be found in Ref. \cite{pitcher2010}.  

\begin{figure}[h]
\includegraphics[width=\columnwidth, trim=0cm 2cm 0cm 2cm]{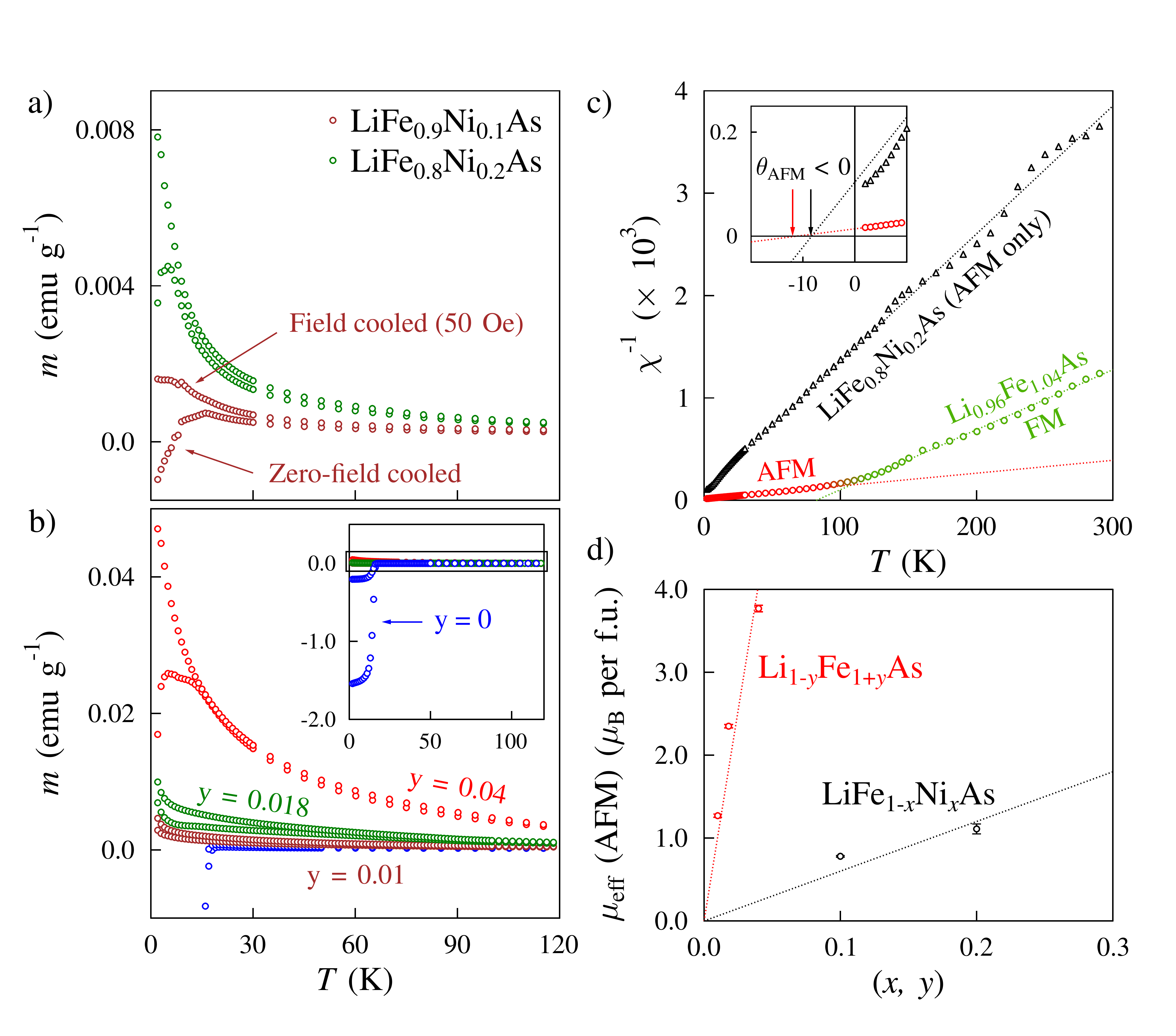}
\caption{Susceptibility data (zero-field cooled and field-cooled in $50\,{\rm Oe}$) and analysis for all samples. The data for a) the LiFe$_{1-x}$Ni$_{x}$As series and b) the Li$_{1-y}$Fe$_{1+y}$As series are presented, with the superconducting response of stoichiometric LiFeAs shown in the inset to (b). Panel (c) compares the inverse susceptibility for LiFe$_{0.8}$Ni$_{0.2}$As and Li$_{0.96}$Fe$_{1.04}$As; the former is demonstrates only antiferromagnetic behaviour [see inset to (c)], whereas correlations in the latter seem to cross over from antiferromagnetic (AFM) to ferromagnetic (FM) on warming. A comparison of the variation in moment size with both $x$ and $y$ is given in (d); the lines through the points are a guide to the eye. All values for calculated moment sizes are given in Table~\ref{mu_eff}.}
\label{squid} 
\end{figure}

Figures \ref{squid}(a) and (b) show the magnetic susceptibility data for the Ni-doped and Fe-rich series respectively, and at first sight they seem similar: both series produce a paramagnetic response and a divergence of field-cooled and zero-field cooled signals at low temperatures, which may indicate a spin-glass transition. Significant differences are revealed, however, upon fitting to a Curie-Weiss dependence [$\chi = C/(T-\theta)$]. Shown most clearly in the inverse susceptibility plots of Fig.\,\ref{squid}(c), it is found that the Fe-rich Li$_{0.96}$Fe$_{1.04}$As sample exhibits both ferromagnetic ($\theta>0$) and antiferromagnetic ($\theta<0$) correlations, whereas only antiferromagnetic behaviour is found in the Ni-doped sample LiFe$_{0.8}$Ni$_{0.2}$As. Additionally, the size of these moments is generally larger across the Fe-rich series compared to the Ni-doped series, as Fig.\,\ref{squid}(d) shows. The values for all extracted moment sizes are given in Table~\ref{mu_eff}.  

\begin{table}[h]
\begin{tabular}{|c|c|c|c|c|}
\hline & \multicolumn{2}{|c|}{AFM} & \multicolumn{2}{|c|}{FM} \\ \cline{2-5}
{\bf Doping} & $\theta$\,(K) & $\mu_{\rm eff}$\,($\mu_{\rm B}$\,/\,f.u.) & $\theta$\,(K) & $\mu_{\rm eff}$\,($\mu_{\rm B}$\,/\,f.u.) \\ 
\hline
\hline Fe$_{1.01}$ & $-21\,(1)$ & $1.27\,(2)$ & - & - \\ 
\hline Fe$_{1.018}$ & $-32\,(2)$ & $2.35\,(2)$ & $30\,(2)$ & $1.47\,(2)$ \\ 
\hline Fe$_{1.04}$ & $-12\,(1)$ & $3.77\,(4)$ & $78\,(4)$ & $1.75\,(5)$ \\ 
\hline Ni$_{0.1}$ & $-9.3\,(5)$ & $0.78\,(1)$ & - & - \\ 
\hline Ni$_{0.2}$ & $-4.7\,(10)$ & $1.11\,(6)$ & - & - \\ 
\hline 
\end{tabular}
\caption{Comparison of effective moment sizes and nature of correlations observed in all samples studied.} 
\label{mu_eff}
\end{table}

In the Li$_{1-y}$Fe$_{1+y}$As series, one might expect these moments to be associated solely with the Fe ions sitting on the Li site, which may act as impurity spins like those in dilute alloys. However, such a scheme cannot account for several features of our data; namely, the sizes of the moments, the fact that these change with Fe concentration and their unusual correlations. Nickel is widely believed to act primarily as an electron donor \cite{pitcher2010, parker2010}, rather than an impurity scatterer, and as such is driving the destruction of superconductivity in LiFe$_{1-x}$Ni$_{x}$As by altering the band filling. We therefore suggest that, as for the Ni-doped series, the moments present in Li$_{1-y}$Fe$_{1+y}$As have an itinerant character. 

To help illuminate the low-temperature phases in these systems, be they glassy or otherwise, we used muon-spin rotation ($\mu$SR). This technique uses the asymmetric emission of positrons during muon decay to track the depolarisation of muons implanted within a sample, thus probing the local field distribution. The details of such experiments can be found in Ref. \cite{blundell1999}. This technique has been useful in studying the rich variety of magnetic states in Fe-based superconductors \cite{klaussLaFeAsO, bakerSrFeAsF, aczelBaFe2As2, jescheSrFe2As2}, particularly when the moment sizes are too small to be detected by other techniques \cite{parker2010, wright2012}. 

\begin{figure*}[t]
\centering
\includegraphics[width=18cm, trim=0cm 2cm 0cm 0cm]{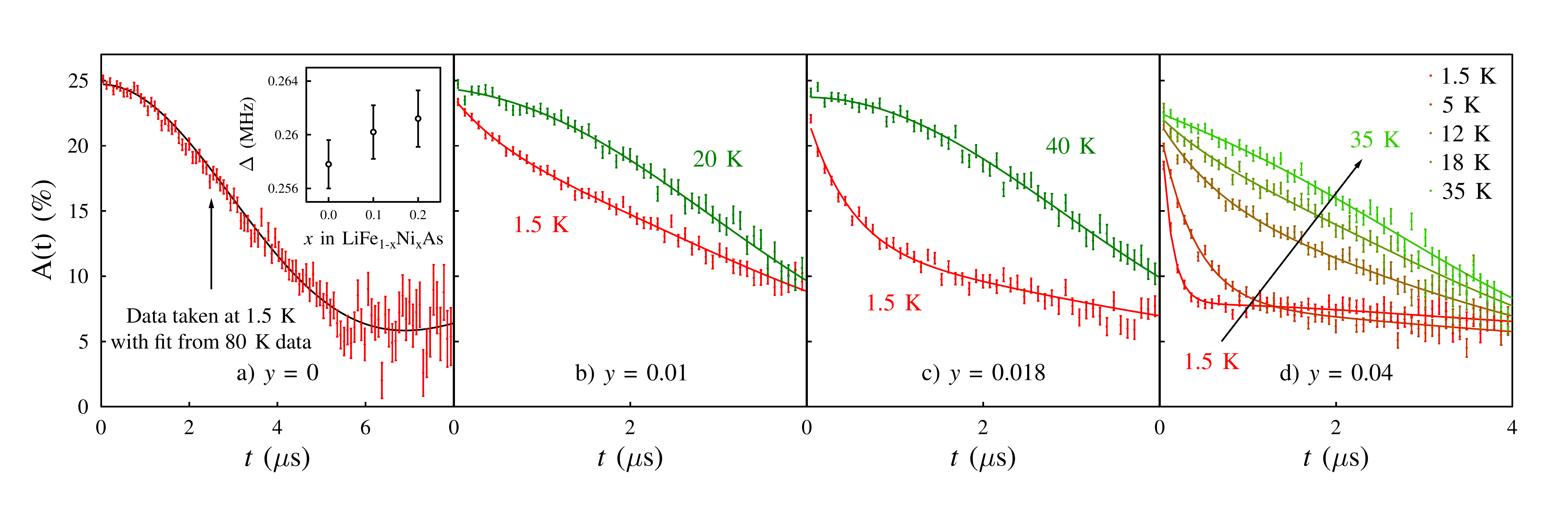}
\caption{A comparison of the zero-field muon data for both the LiFe$_{1-x}$Ni$_x$As and Li$_{1-y}$Fe$_{1+y}$As compounds. The data for stoichiometric LiFeAs, taken at $1.5\,{\rm K}$, are shown in panel (a). The superimposed fit is taken from the $80\,{\rm K}$ data, emphasising that no temperature-dependent changes were observed. The data for the Ni-doped series (not shown) can be similarly described by a single temperature-independent Kubo-Toyabe function, albeit with widths ($\Delta$) that increase slightly with $x$ ({\it Inset}). For $y\geq\,0.01$ in the Li$_{1-y}$Fe$_{1+y}$As series [panels (b)--(d)], a Kubo-Toyabe function accounts for high-temperature data, but an exponential relaxation emerges on cooling that is more pronounced for samples with a higher Fe concentration.}
\label{compare} 
\end{figure*}

Figure \ref{compare} summarises the data for all samples, taken in zero applied field (ZF). For stoichiometric LiFeAs [Fig.\,\ref{compare}(a)] the data are best described by a single Gaussian Kubo-Toyabe function, suggesting that the muons experience a field solely due to randomly orientated, quasistatic nuclear dipole moments \cite{blundell1999}. The fit remains unaltered across the entire temperature range; Fig.\,\ref{compare}(a) shows the data taken at $1.5\,{\rm K}$ with the fit from $80\,{\rm K}$ superimposed. Together with the observation of antiferromagnetic fluctuations in INS experiments \cite{taylor2011, qureshi2012, wang2011}, our data cast doubt on the triplet-pairing predictions in Refs. \cite{brydon2011, aperis2013}. Zero-field (ZF-) $\mu$SR is known to be sensitive to the small magnetic fields induced under spontaneous time-reversal symmetry breaking (which may be due to triplet pairing) in Sr$_2$RuO$_4$ \cite{luke1998} and LaNiC$_2$ \cite{hillier2009}. On crossing $T_{\rm c}$ no such fields can be resolved here, so we find no evidence to support a triplet pairing hypothesis in LiFeAs.

We find that the $\mu$SR data for our two members of the LiFe$_{1-x}$Ni$_x$As series can also be described by temperature-independent Kubo-Toyabe functions which are almost identical to that which describes the LiFeAs data. The only difference is a slight increase of the second moment of the local magnetic field distribution, $\Delta$, observed as the Ni concentration, $x$, increases [see Fig.\,\ref{compare}(a)({\it Inset})]. This is a consequence of the Ni nuclear moment ($-0.75\,\mu_{\rm N}$) being significantly larger than that of Fe ($0.09\,\mu_{\rm N}$). These data suggest that the LiFe$_{1-x}$Ni$_x$As samples exhibit a dynamically fluctuating state at all measured temperatures and do not exhibit a spin-frozen state, despite some apparently glassy behaviour observed in the susceptibility data. 

By contrast, an emergent magnetic phase is identified in the Li$_{1-y}$Fe$_{1+y}$As series: Figure \ref{compare}(b), (c) and (d) show the ZF data for the $y\,=\,0.01, 0.018$ and $0.04$ members of the Li$_{1-y}$Fe$_{1+y}$As series respectively. In all three samples we observe Kubo-Toyabe functions at high temperatures, but the relaxation becomes more exponential on cooling. These data are consistent with a freezing of the induced moments and the effect is clearly stronger for samples with higher Fe concentrations. 

To analyse the $\mu$SR data, we assumed the existence of two distinct muon sites in the unit cell, as found in related materials \cite{maeter2009}. In the ordered phases of NaFeAs, one observes two frequencies related by a constant factor of $\sim 10$ which indicates the relative coupling strengths between each of these two muon sites and the ordered Fe moments. For our isostructural Li$_{1-y}$Fe$_{1+y}$As series, we therefore fitted our spectra to the two-component function   

\begin{figure}[h]
\includegraphics[width=\columnwidth, trim=0cm 2cm 0cm 2cm]{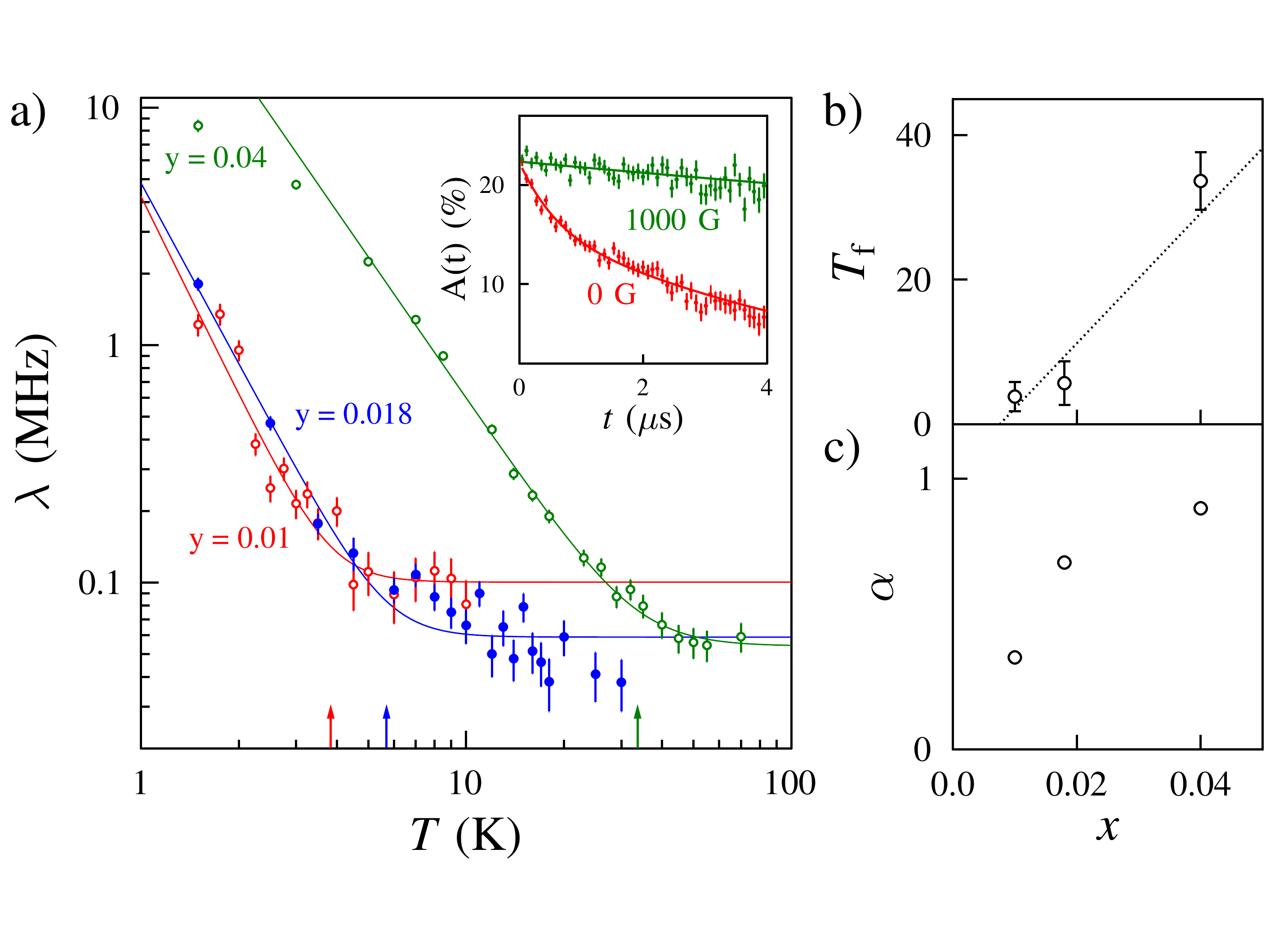}
\caption{(a) The evolution of $\lambda(T)$ for the Li$_{1-y}$Fe$_{1+y}$As series. The spin-freezing temperature, $T_{\rm f}$, is defined as the onset of the power law increase in $\lambda(T)$. {\it Inset:} spectra at $10\,{\rm K}$ in both zero-field and a longitudinal field (LF) of $1000\,{\rm G}$. The weak relaxation still present in the LF spectrum indicates dynamic behaviour, pointing to a spin-freezing picture, as opposed to static local order (see text). (b) The values of the $T_{\rm f}$ extracted from the behaviour of $\lambda(T)$. (c) The variation of the fast relaxing amplitude, $\alpha$ (defined in Eq.\,\ref{asym}), with Fe concentration. \label{zf_full}}
\end{figure}

\vspace{-0.7cm}
\begin{center}
\begin{equation}
A(t) = G_{\rm KT}(\Delta, t)\,[\alpha\,{\rm e}^{-\lambda t}+(1-\alpha)\,{\rm e}^{-\lambda Rt}],
\label{asym}
\end{equation}
\end{center}

\noindent
where the relaxation rate ratio, $R$, was fixed throughout the fitting. The fractional amplitude of each contribution, $\alpha$, did not vary with temperature but did scale with $y$: going from $0.35$ for $y\,=\,0.01$ to $0.9$ for $y\,=\,0.04$ [this is shown in Fig.\,\ref{zf_full}(c)]. The free parameter, $\lambda$, is the relaxation rate describing fluctuations in the local magnetic field caused by the dynamics of the electronic moments. Both muon sites will experience relaxation due to nuclear moments and so a Gaussian Kubo-Toyabe function, with a fixed width ($\Delta$), was included as an overall multiplicative component.   

\begin{figure}[h]
\includegraphics[width=\columnwidth, trim=0cm 1cm 0cm 1cm]{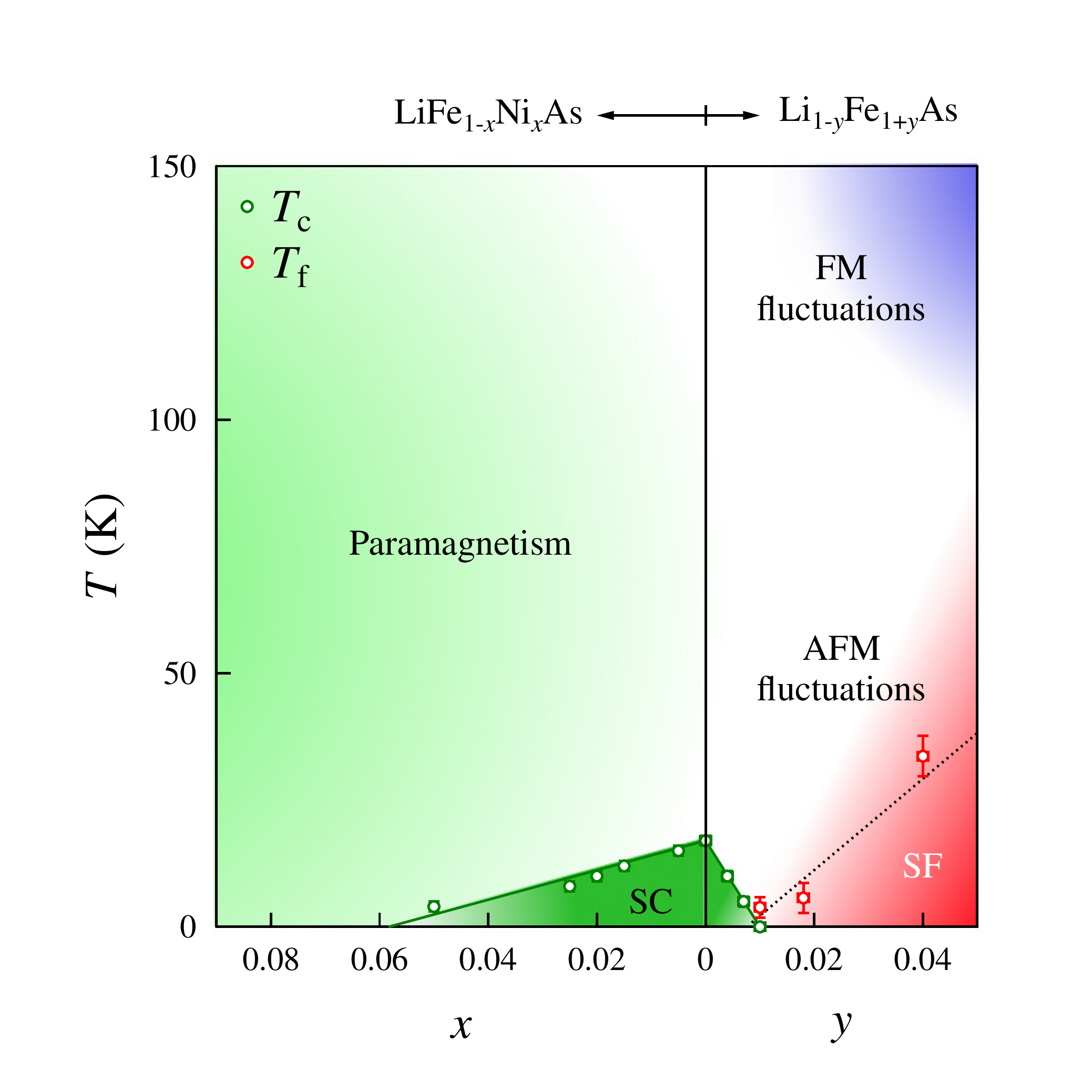}
\caption{Phase diagram for the Li$_{1-y}$Fe$_{1+y}$As and LiFe$_{1-x}$Ni$_x$As series, showing regions of superconductivity (SC), spin freezing (SF) as well as (anti)ferromagnetic fluctuations. \label{phase}}
\end{figure}

Figure~\ref{zf_full}(a) plots the temperature variation of the larger relaxation rate $\lambda$ for all samples, along with a comparison of spectra taken in zero-field and a longitudinal field (LF) of $1000\,{\rm G}$ for Li$_{0.96}$Fe$_{1.04}$As at $10\,{\rm K}$. The weak relaxation observed in the LF spectra suggests that these moments are dynamically fluctuating, and that the increase in $\lambda$ seen at low temperatures corresponds to a slowing-down of these fluctuations. The best fit lines in Fig.\,\ref{zf_full}(a) assume that both power-law and temperature-independent relaxation processes contribute to $\lambda(T)$ in quadrature. We can define the spin freezing temperature, $T_{\rm f}$, for each sample as the onset of the power-law increase in $\lambda(T)$ (defined as the temperature at which the two contributions to $\lambda(T)$ are equal), as shown by the arrows in Fig.\,\ref{zf_full}(a). These values are plotted in Fig.\,\ref{zf_full}(b) and are used to compose the phase diagram in Fig.\,\ref{phase}.

It was difficult to resolve the second (smaller) relaxation rate, so in fact the plots of $\lambda(T)$ in Fig.\,\ref{zf_full}(a) were obtained with the ratio $R$ set to zero. This is unsurprising if we can assume the observed dynamics operate in the fast-fluctation limit, such that $\lambda = 2\Delta_{\rm site}^2/\nu$; where $\Delta_{\rm site}/\gamma_{\mu}$ is the rms value of the local field at a given muon site, and $\nu$ is the fluctuation rate. Because the strength of the dipolar coupling to moments on the Fe site probably differs between the two muon sites by a factor of $\sim 10$ (based on the frequencies seen in NaFeAs), we would expect the relative relaxation rates to differ by a factor of $\sim 100$. This explains our inability to fit the smaller relaxation rate explicitly.

An alternative explanation would be to assume a single relaxation rate and interpret Eq.\,\ref{asym} as describing mesoscopic phase separation where a fraction of muons sit in isolated regions of local order. However, the dynamic behaviour observed in LF spectra, the glassy behaviour seen in the susceptibility data and the overwhelming evidence for a two-site model from isostructural systems leads us to suggest the picture outlined above is the most plausible explanation for what we observe.       

In conclusion, we have demonstrated that suppressing superconductivity by subsituting Fe onto the Li site induces fluctuating, correlated, itinerant moments that freeze at low temperatures. On warming, the SQUID data show that magnetic correlations are most likely antiferromagnetic in nature, but there appears to be a cross-over to ferromagnetic correlations for samples with the largest Fe concentration (see Fig.\,\ref{phase}). The changing size of these moments and the strength of their correlation cannot be explained as simply being the result of incorporating dilute Fe moments onto the Li site, and demonstrates emergent itinerant magnetic behaviour. The effects of an induced moment were proposed in relation to an anomalous result from earlier work on a supposedly Li-deficient sample \cite{pratt2009}; we now believe that this sample may have also contained a small amount of Fe on the Li site ($y\,<\,0.01$) which could result in a similar spin freezing effect. No evidence of any such behaviour is observed if superconductivity is suppressed by Ni subsitution onto the Fe site, so this state is not associated with a general suppression of superconductivity but is unique to samples with Fe substituted for Li. Despite a detailed search, we find no evidence of spontaneous fields below $T_{\rm c}$ in stoichiometric LiFeAs and thus no evidence of triplet pairing. 

Part of this work was carried out using the GPS Spectrometer at the Paul Scherrer Institut, Villigen, CH. We acknowledge the financial support of EPSRC (UK).

\end{document}